\documentstyle[12pt]{article}

\catcode`\@=11 \@addtoreset{equation}{section}\catcode`\@=12

\begin{document}
\thispagestyle{empty}

\begin{center}
               RUSSIAN GRAVITATIONAL ASSOCIATION\\
               CENTER FOR SURFACE AND VACUUM RESEARCH\\
               DEPARTMENT OF FUNDAMENTAL INTERACTIONS AND METROLOGY\\
\end{center}
\vskip 4ex
\begin{flushright}                              RGA-CSVR-004/94\\
                                                gr-qc/9406050
\end{flushright}
\vskip 15mm

\begin{center}
{\large\bf
Multitemporal generalization \\
of Schwarzschild solution }

\vskip 5mm
{\bf
Vladimir D. Ivashchuk and Vitaly N. Melnikov }\\
\vskip 5mm
     {\em Centre for Surface and Vacuum Research,\\
     8 Kravchenko str., Moscow, 117331, Russia}\\
     e-mail: mel@cvsi.uucp.free.msk.su\\
\end{center}
\vskip 10mm
ABSTRACT

The  $n$-time  generalization  of  Schwarzschild  solution   is
presented.  The   equations  of  geodesics  for  the  metric  are
integrated and the  motion of the relativistic particle is considered.
The multitemporal analogue of the Newton's gravitational law for the
objects, described by the solution, is suggested. The scalar-vacuum
generalization of the  multitemporal solution is also presented.

\vskip 10mm


\vskip 30mm

\centerline{Moscow 1994}
\pagebreak

\setcounter{page}{1}

\pagebreak

\section{Introduction}

In [1]  the generalization  of the  Schwarzschild solution to
the  case  of $n$  internal  Ricci-flat  spaces  was obtained.(The
case $n =1$ was considered earlier in [2].) In [3] this  solution  was
generalized  on  $O(d+1)$-symmetric (Tangherlini-like) case. (In [4] the
special case of the solution [3] with $n =2$ was considered).

This paper is devoted to an interesting special case of  the
solution [1].  This is the  $n$-time  generalization  of   the
Schwarzschild solution.  We note,  that  the idea  of considering of
space-time manifolds  with extra time directions   was  discussed  earlier
by different authors  (see, for example, [5-12]).   Some  revival  of  the
interest in  this direction was inspired recently by string models
[9-12].

In sec. 2  the multitemporal generalization of Schwarzschild formula
is considered and corresponding geodesic equations are integrated.
In sec. 3 the motion of the relativistic particle in the background
of the solution is investigated  and  a multitemporal analogue
of the Newton's formula is obtained. The  sec. 4 is devoted to
multitemporal generalization of Newton's mechanics and Newton's
gravitational law for interacting objects described by the solution
("multitemporal hedgehogs").

\section{The metric and geodesic equations}

The  metric  generalizing   the  Schwarzschild  solution  to  the
multitemporal case reads
\begin{equation}
g= - \sum_{i=1}^{n} f^{a_{i}} dt^{i} \otimes dt^{i} +
f^{-b}  dR \otimes dR +
f^{1 - b} R^{2} d \Omega^{2},
\end{equation}
where $f = 1 - (L/R)$, $L = const$, $d \Omega^{2}$ is standard metric on
2-dimensional sphere and the parameters $b, a_{i}$ satisfy
the relations
\begin{equation}
b = \sum_{i=1}^{n} a_{i}, \qquad
b^{2} + \sum_{i=1}^{n} a_{i}^{2} =2.
\end{equation}
The metric (2.1) satisfies the Einstein equations (or
equivalently ${R_{MN}}[g] = 0$) and may be obtained as a
special case of the solution [1] or more general solution [3].

The metrics ${g}(a,L)$ and ${g}(-a,-L)$ are equivalent for any set
$a = (a_1, \ldots , a_n)$, satisfying (2.2). This may be easily
verified using the following transformation of the radial variable:
$R = R_{*} + L$. So, without loss of generality we restrict our
consideration by the case $L > 0$ (the case $L = 0$ is trivial).

In the case
\begin{equation}
a_{i} = \delta_{ik},
\end{equation}
$k \in \{1,...,n \}$, the metric (2.1) has the following form
\begin{equation}
g=g^{(k)}_{Sch} - \sum_{i \neq k } dt^{i} \otimes dt^{i} ,
\end{equation}
i.e. it is a trivial (cylindrical)
extension of the Schwarzschild solution  with the time $t^{k}$.
It describes an extended (in times) membrane-like object.
Any section of this object by hypersurface
$t^{i} = t^{i}_{0} =const$, $i \neq k$, is the $4$-dimensional
black hole, "living" in the time $t^{k}$. It may be proved
that the solution (2.1) has a singularity at $R = L$
for all sets of parameters $(a_1, \ldots , a_n)$ except
$n$ Schwarzschild-like points (2.3) (for $n=2$ this was proved
in [14]).

We consider the geodesic  equations for the metric (2.1)
\begin{equation}
\ddot{x}^{M} + {\Gamma^{M}_{NP}}[g]\dot{x}^{N}\dot{x}^{P} = 0,
\end{equation}
where  $x^{M} = {x^{M}}(\tau)$, $\dot{x}^{M} = dx^{M}/d\tau$
and $\tau$ is some parameter on a curve.

These equations are nothing
more than the Lagrange  equations for the Lagrangian
\begin{eqnarray} L_{1} &&= \frac{1}{2}{g_{MN}}(x) \dot{x}^{M}\dot{x}^{N}
  \nonumber  \\
  &&= \frac{1}{2} [ f^{-b}(\dot{R})^{2} + f^{1 - b} R^{2}
  (\dot{\theta}^{2} + \sin^{2}\theta \dot{\varphi}^{2}) - \sum_{i=1}^{n}
  f^{a_{i}}(\dot{t}^{i})^{2}].
\end{eqnarray}

The complete set of integrals of motion for the Lagrange system (2.6)
is following
\begin{eqnarray}
&&f^{a_{i}} \dot{t}^{i} = \varepsilon^{i},  \\
&&f^{1-b}R^{2} \dot{\varphi} = j,  \\
&& f^{-b}\dot{R}^{2} + f^{1 - b} R^{2} \dot{\varphi}^{2}
- \sum_{i=1}^{n} f^{a_{i}}(\dot{t}^{i})^{2} = 2E = 2L_1,
\end{eqnarray}
$i = 1, ..., n$. Without loss of generality we put
here $\theta = \frac{\pi}{2}$.

{\bf Multitemporal horizon.} Here we  consider the null geodesics.
Putting $E =0$ in (2.9) we get for a light "moving" to the center
\begin{equation}
\dot{R} = - \sqrt{\sum_{i=1}^{n} (\varepsilon^{i})^{2} f^{b-a_{i}}
 - j^{2} f^{-1+2b} R^{-2}}
\end{equation} and consequently
\begin{equation} t^{i} -
t_{0}^{i} = - \int_{R_{0}}^{R} dx \frac{\varepsilon^{i}
[{f}(x)]^{-a_{i}}} {\sqrt{\sum_{i=1}^{n} (\varepsilon^{i})^{2}
[{f}(x)]^{b-a_{i}} - j^{2} [{f}(x)]^{-1+2b} x^{-2}}} ,
\end{equation}
$i = 1, \ldots ,n$.

Let $L > 0$, $\varepsilon = (\varepsilon^{i}) \neq 0$
and $a = (a_1, \ldots , a_n)$ satisfies (2.2). We say that
the $\varepsilon$-horizon takes place for the metric (2.1) at $R =
L $ if and only if
\begin{equation} ||t - t_{0}|| \equiv
\sum_{i=1}^{n} |t^{i} - t_{0}^{i}| \rightarrow + \infty, \end{equation} as
$R \rightarrow L$ for all $t_{0}$ and $j$. It may be proved [14]
that  for $L > 0$ and for non-Schwarzschild set $a$ the
$\varepsilon$-horizon for the metric (2.1) at $R = L$ is absent
for any $\varepsilon \neq 0$. For the Schwarzschild set of parameters
(2.3) the $\varepsilon$-horizon takes place if $\varepsilon^{k} \neq 0$,
i.e. light should "move" in $t^k$-direction.

\section{Relativistic particle}

Let us consider the motion of the relativistic particle in
the gravitational  field, corresponding  to the  metric (2.1). The
Lagrangian of the particle is
\begin{equation}
L_{2} = -m \sqrt{-{g_{MN}}(x) \dot{x}^{M} \dot{x}^{N}},
\end{equation}
where $m$ is the mass of the particle.
     The Lagrange equations for (3.1) in the proper time gauge
\begin{equation}
{g_{MN}}(x) \dot{x}^{M} \dot{x}^{N} = -1
\end{equation}
coincide with the geodesic equations (2.5). In this case
$(E^{i})= (m \varepsilon^{i})$ is the energy vector  and $J = mj$ is
the angular momentum.  For fixed values of $\varepsilon^{i}$ the
3-dimensional part of the equations of motion is generated by the Lagrangian
\begin{equation}
L_{*} = \frac{m}{2}[f^{1-b} \bar{g}_{Sch, \alpha \beta}(x)
\dot{x}^{\alpha}\dot{x}^{\beta}
+ \sum_{i=1}^{n} (\varepsilon^{i})^{2} f^{-a_{i}}].
\end{equation}
where $\bar{g}_{Sch}$ is the space section of the Schwarzschild metric.

Now, we restrict our consideration by the non-relativistic
motion at large distances: $R \gg L$. In this approximation:
$t^{i} = \varepsilon^{i} \tau, \qquad \sum_{i=1}^{n}
(\varepsilon^{i})^{2} = 1.$
It follows from (3.3) that in the considered approximation we get a
non-relativistic particle of mass $m$, moving in the  potential
\begin{equation}
V = - \frac{m}{2} \sum_{i=1}^{n} (\varepsilon^{i})^{2} \frac{a_{i}L}{R}
= - G \frac{m(\varepsilon^{i} M_{ij} \varepsilon^{j})}{R},
\end{equation}
where $G$ is the gravitational constant and
\begin{equation}
M_{ij} = a_{i} \delta_{ij} L/ 2G,
\end{equation}
are the components of the gravitational mass matrix.

We note, that  the relation (3.4) may be rewritten as following
\begin{equation}
V = -G \frac{ tr(M M_{I})}{R}
\end{equation}
where  $M_{I} = (m \varepsilon_{i}\varepsilon_{j})$ is the inertial
mass matrix of the particle. For $n = 1$  the potential (3.6) coincides
with the Newton's one.

{\bf Matrix form.} The  solution (2.1) may be also rewritten in the matrix
form
\begin{eqnarray}
g=&& -[(1-L/R)^A]_{ij} d \bar{t}^{i} \otimes d \bar{t}^{j} \nonumber \\
&& + (1 - L/ R)^{- trA}  dR \otimes dR + (1 -
L/R)^{1 - trA} R^{2} d \Omega^{2},
\end{eqnarray}
where $A$ is a real
symmetric $n \times n$-matrix satisfying the relation 
\begin{equation}
(tr A)^{2}+ tr (A^{2}) = 2.
\end{equation}
Here $x^{A} \equiv  \exp (A \ln x)$ for $x > 0$. The metric (3.7) can
be reduced to the metric (2.1) by the diagonalization of the $A$-matrix:
$A= S^{T}(a_{i} \delta_{ij}) S$, $S^{T} S = 1_{n}$ and
the reparametrization of the time variables: $S^{j}_i
\bar{t}^{i} =  t^{j}$.
In this case the gravitational mass matrix is
\begin{equation}
(M_{ij}) = (A_{ij} L/ 2G).
\end{equation}
We may also define the gravitational mass tensor as
\begin{equation}
{\cal M} = M_{ij} d \bar{t}^{i} \otimes d \bar{t}^{j}.
\end{equation}

We call the extended (in time)
object, corresponding to the solution
(3.7)-(3.8) as multitemporal Schwarzschild hedgehog. At large distances
$R \gg L$ this object
is described by the matrix analogue of the Newton's
potential
\begin{equation}
\Phi_{ij} = - \frac{1}{2} L A_{ij}/R = -G  M_{ij}/R.
\end{equation}
Clearly, that this potential for the diagonal case (2.1)
$A = a_{i} \delta_{ij}$ is a superposition of the potentials,
corresponding to "pure states": Schwarzschild-like
membranes (2.4).  So, in the
post-Newtonian approximation the Schwarzschild hedgehog
is equivalent to the superposition of black hole membranes (2.4),
corresponding to different times.

\section{Multitemporal Newton laws}

The solution (3.7), (3.8)  may be also rewritten
as following
\begin{eqnarray}
g=&& -[(1- ||L||/R)^{L/||L||}]_{ij} d t^{i} \otimes d
t^{j} \nonumber \\
&&+ (1 - ||L||/ R)^{- tr L/||L||}  dR \otimes dR +
(1 - ||L||/R)^{1 - (tr L/||L||)} R^{2} d \Omega^{2},
\end{eqnarray}
where here $L = (L_{ij}) \neq 0$ is  real
symmetric $n \times n$-matrix  with the norm
\begin{equation}
||L|| \equiv \sqrt{\frac{1}{2}(tr L)^{2}+ \frac{1}{2} tr (L^{2})}.
\end{equation}
We call matrix $L$ as gravitational length matrix.

Now we consider the interaction between two multitemporal hedgehogs
with gravitational length matrices $L_1= (L_{1,ij})$ and $L_2 = (L_{2,ij})$
located at large distances from each other
\begin{equation}
|\vec{x}| \gg  ||L||_1, ||L||_2, \qquad \vec{x} \equiv
 \vec{x}_1 -  \vec{x}_2.
\end{equation}
We begin with the simplest case $n=1$.
In  Newton's mechanics the equations of motion
for two point-like masses $M_1 = L_1/2G$ and  $M_2 = L_2/2G$
with world lines $\vec{x}_1 = {\vec{x}_1}(t)$  and
$\vec{x}_2 = {\vec{x}_2}(t)$  respectively are well-known:
\begin{eqnarray}
&& \frac{d^2 \vec{x}_1}{dt^2} = - L_2 \frac{\vec{x}}{2|\vec{x}|^3}, \\
&& \frac{d^2 \vec{x}_2}{dt^2} =  L_1 \frac{\vec{x}}{2|\vec{x}|^3},
\end{eqnarray}
where $\vec{x}$ is defined in (4.3). The equations (4.4), (4.5)
may be obtained from the Einstein equations, when the solutions
describing  the post-Newtonian (4.3), non-relativistic motion
\begin{equation}
|\frac{d \vec{x}_a}{dt}|  \ll 1,
\end{equation}
$a =1,2$, of two black holes are considered.

Our hypothesis is that the  generalization of
this scheme to the multitemporal case should
lead to the following equations of motion for two
non-relativistic hedgehogs with gravitational length matrices
$L_1$ and $L_2$
in the post-Newtonian approximation (4.3)
\begin{eqnarray}
&& \frac{d^2 \vec{x}_1}{dt^i dt^j}
= - L_{2,ij} \frac{\vec{x}}{2|\vec{x}|^3}, \\
&& \frac{d^2 \vec{x}_2}{dt^i dt^j} =  L_{1,ij} \frac{\vec{x}}{2|\vec{x}|^3}.
\end{eqnarray}
The functions $\vec{x}_a = {\vec{x}_a}(t_1, \ldots, t_n)$, $a=1,2$,
describe the world surfaces of two multitemporal
objects in the considered approximation. The multitemporal analogue of
the non-relativistic condition  (4.6) reads
\begin{equation}
|\frac{d \vec{x}_a}{dt^i}|  \ll 1,
\end{equation}
$a=1,2$, $i=1, \ldots, n$. Defining gravitational mass matricies
\begin{equation}
(M_{a,ij}) = (L_{a,ij} / 2G),
\end{equation}
and forces
\begin{equation}
\vec{F}_{a,ij} = M_{a,ij} \frac{d^2 \vec{x}_a}{dt^i dt^j},
\end{equation}
$a =1, 2$, we get
\begin{eqnarray}
&&\vec{F}_{1,ij} = - G M_{1,ij} M_{2,ij} \frac{\vec{x}}{|\vec{x}|^3}, \\
&&\vec{F}_{1,ij} = - \vec{F}_{2,ij},
\end{eqnarray}
$i,j =1, \ldots ,n$. Relations (4.11), (4.12) and (4.13) are
multitemporal analogues of the Newton's laws ,
describing the multitemporal "motion" of  two interacting
non-relativistic hedgehogs in the post-Newtonian approximation.
(The generalization to multi-hedgehog case is quite transparent.)
We note, that for  $\vec{F}_{1} = tr (\vec{F}_{1,ij})$ we get
the formula suggested previously in [15]
\begin{equation}
\vec{F}_{1} = - G tr (M_{1} M_{2}) \frac{\vec{x}}{|\vec{x}|^3}.
\end{equation}

{\bf Scalar-vacuum generalization.} The solution (2.1) can be easily
generalized on scalar-vacuum case. In this case the field equations
corresponding  to the action
\begin{equation}
S = \frac{1}{2}
\int d^{D}x \sqrt{|g|} ({R}[g] -
\partial_{M} \varphi \partial_{N} \varphi g^{MN}),
\end{equation}
are satisfied for the metric (2.1) and the scalar field
\begin{equation}
 \varphi = \frac{1}{2} q \ln (1 - \frac{L}{R}) + const,
\end{equation}
with the parameters related as following
\begin{equation}
b = \sum_{i=1}^{n} a_{i}, \qquad
b^{2} + \sum_{i=1}^{n} a_{i}^{2} + q^2  = 2.
\end{equation}
This solution is a special case of the solution [16] or more
general dilatonic-electro-vacuum solution [14,17].

\begin{center} {\bf Conclusion}   \end{center}

In this paper we considered the multitemporal generalization
of the Schwarzschild solution. We integrated the equations of geodesics
for the metric and considered the motion of relativistic particle in the
background , corresponding to the metric. We obtained the modification
of Newton's law for interaction of massive non-relativistic particle
with multitemporal hedgehog (i.e extended in time object, described
by the solution). We also suggested multitemporal analogues of Newton's
formulas for non-relativistic motion of interacting hedgehogs. We note,
that the main difference of the multitemporal ($n$-time) case from the
ordinary $n=1$ case is following:  in the space-time with $n$ time
coordinates the gravitational and inertial masses are $n \times n$
matrices, and the energy of a relativistic particle is the
$n$-component vector.

%


\end{document}